\documentclass[reprint,
amsmath,amssymb,
aps,
prb,
superscriptaddress,
floatfix]{revtex4-2}

\usepackage[x11names, svgnames, dvipsnames]{xcolor}
\definecolor{maroon}{RGB}{128,0,0}
\usepackage{hyperref}
\hypersetup{colorlinks,linkcolor={maroon},citecolor={maroon},urlcolor={black}}
\usepackage{cleveref}
\usepackage{graphicx}
\usepackage{float} % for [H] figure placement
\usepackage{dcolumn}
\usepackage{bm}
\usepackage{comment}
\usepackage{braket}
\usepackage{textcomp, gensymb}
\usepackage{subcaption}

\crefname{figure}{Fig}{Figs}

\begin{document}

\preprint{APS/123-QED}

\title{Quasi-long-range order in the quantum XY model on Sierpi\'nski gaskets}

\author{Robert Canyellas}
\affiliation{Institute for Molecules and Materials, Radboud University,
Heyendaalseweg 135, 6525AJ Nijmegen, The Netherlands}

\author{Jannes van Poppelen}
\affiliation{Department of Physics and Astronomy, Uppsala University,
Box 516, SE-75120 Uppsala, Sweden}

\author{Annica M. Black-Schaffer}
\affiliation{Department of Physics and Astronomy, Uppsala University,
Box 516, SE-75120 Uppsala, Sweden}

\author{Mikhail I. Katsnelson}
\affiliation{Institute for Molecules and Materials, Radboud University,
Heyendaalseweg 135, 6525AJ Nijmegen, The Netherlands}

\author{Andrey A. Bagrov}
\affiliation{Institute for Molecules and Materials, Radboud University,
Heyendaalseweg 135, 6525AJ Nijmegen, The Netherlands}

\date{\today}

\begin{abstract}
Fractal geometry has recently been shown to enhance the amplitude of the
superconducting order parameter, but the fate of the superconducting phase on such lattices is far
less clear. Classical XY models on finitely ramified fractals have no finite-temperature
transition, as the energy of a vortex stays bounded as the system grows. However, that
statement concerns an infinite gasket and thermal vortices, while real fractal
structures are finite by construction. It also leaves open what happens at $T = 0$,
where the phase competes not with temperature but with its own quantum
fluctuations. In this work we address whether superconducting phase coherence survives at $T=0$. 
To that end, we study the
spin-$1/2$ XY ferromagnet, the phase sector of a Josephson network with at most one
Cooper pair per island, thus capturing the maximum effect of quantum phase fluctuations, on both triangular and honeycomb Sierpi\'nski gaskets. By using
vision-transformer neural quantum states optimized by variational Monte Carlo, we
reach generations up to $G = 5$, that is $366$ sites for the triangular gasket and
$609$ for the honeycomb one, well beyond the reach of exact diagonalization. We find that the
transverse two-point correlator decays algebraically, $C_{xy}(r) \sim r^{-\eta}$
with $\eta \simeq 0.12\!-\!0.32$, independent of distance measured along the graph, along
the fractal boundary, or in the embedding plane. Quantum phase fluctuations thus
leave quasi-long-range order intact in the ground state. We further discuss
the implication of these results for the superconducting response of a finite gasket at low but nonzero temperature. 
\end{abstract}

\maketitle

\section{Introduction}

Quantum dynamics on artificial fractal lattices has recently become an experimentally
accessible subject. Fractal arrays of artificial atoms have been assembled by positioning CO molecules on Cu(111)~\cite{kempkes2019design} or grown by molecular
self-assembly~\cite{shang2015assembling}, as well as realized as bismuth nanostructures
hosting topological edge and corner states~\cite{canyellas2024topological}.
Mesoscopic superconducting realizations go back even further: Josephson junction arrays and superconducting wire networks patterned as Sierpi\'nski gaskets have been measured
since the late 1980s~\cite{gordon1986fractal,meyer1991vortex,meyer2002dimensional},
and fractal Josephson weak links still attract
attention~\cite{amundsen2024josephson}.

Recent calculations have also sharpened the case for superconductivity in fractal systems. Mean-field Bogoliubov-de Gennes (BdG) treatment of the attractive Hubbard model on the Sierpi\'nski gasket has been shown to give a superconducting critical temperature well
above that of the regular triangular lattice, with a superfluid stiffness large enough to support macroscopic phase coherence \cite{iliasov2026strong, canyellas2026extended}. The Sierpi\'nski carpet shows no such
effect, a difference attributed to the ramification properties of the two fractals~\cite{iliasov2026strong}. Furthermore, for order parameters of non-trivial symmetry, the fractal geometry reshuffles the competition between pairing channels~\cite{canyellas2026extended}, while related boundary-driven enhancements appear on hyperbolic lattices~\cite{bashmakov2025hyperbolic}. The underlying mechanism is spectral: a finitely ramified fractal compresses
the single-particle eigenvalues into narrow, strongly degenerate clusters, which favors superconducting pairing.

The above results are, however, mean-field statements about the pairing amplitude, while in low dimensions the
superconducting transition temperature is usually controlled not by the superconducting amplitude, but by its phase. On a fractal lattice the phase is the problematic part.
Early results by Vallat, Korshunov, and Beck have shown that the energy of a vortex on a two-dimensional Sierpi\'nski gasket does not
grow with the system size but saturates, such that free vortices are present at arbitrarily
low temperatures and thus no Berezinskii-Kosterlitz-Thouless (BKT) transition into the superconducting state can occur~\cite{vallat1991xy}.

Monte Carlo simulations of the classical XY model have confirmed the lack of a BKT transition for the
Sierpi\'nski gasket~\cite{mitrovic2010monte} and for the 
Sierpi\'nski pyramid~\cite{przedborski2011absence}. The mechanism is the same in both cases: on a
finitely ramified structure an arbitrarily large region can be disconnected by
cutting a finite number of bonds, such that an excitation that destroys coherence costs an
energy that does not grow with the system size, while the entropy of placing it does. This is in fact the argument that forbids ordering in the one-dimensional Ising model, transplanted to a fractal~\cite{gefen1980critical,gefen1984phase}, thus putting the
transition temperature at zero. Interestingly, the Menger sponge, whose order of ramification is
infinite, escapes this reasoning, yet Monte Carlo finds no transition
there either~\cite{przedborski2014absence}. The quantum case has been addressed only for antiferromagnetic exchange, where
exact diagonalization of the $s = 1/2$ model on a gasket consisting of $N = 42$ sites has found a disordered ground state with a very short correlation
length~\cite{voigt2004anisotropic}. The disorder there comes from frustration rather than from the fractal geometry:
antiferromagnetic bonds on the odd loops of the gasket cannot be satisfied simultaneously. The Josephson mapping that we consider here, however, gives the ferromagnetic sign, for which the same triangles are unfrustrated.

The mean-field BdG and classical XY-model results discussed above still leave an important question unsolved. While the mean-field results of the
pairing amplitude are encouraging for superconductivity, the classical results on phase fluctuations
are not, but neither addresses the question of the ground state at zero temperature. That is, on a fractal geometry, does the
superconducting phase have a coherent ground state at all, or do quantum fluctuations disorder it
already at $T = 0$? Note that the classical arguments concern thermal vortex proliferation and say nothing about $T = 0$, where the phase instead competes with the charging energy rather than with temperature. However, if quantum fluctuations destroy off-diagonal order in the ground state, the mean-field enhancement of the gap is of no practical use. 

\begin{figure}
    \centering
    \includegraphics[width=\linewidth]{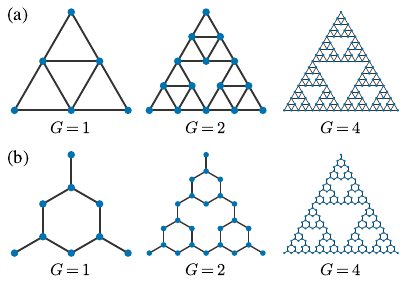}
    \caption{Examples of different generations $G=1,2,4$ for the (a) triangular and (b) honeycomb Sierpi\'nski gaskets.}
    \label{fig:sierpinski}
\end{figure}

Here we address this question numerically. We compute the ground state of the
spin-$1/2$ XY ferromagnet, the extreme quantum limit of a Josephson network with at
most one Cooper pair per island, on several different generations of both the triangular and honeycomb Sierpi\'nski gaskets, see Fig.~\ref{fig:sierpinski}. To achieve this, we use a vision-transformer (ViT) neural quantum state (NQS) optimized with variational Monte Carlo. We find that the transverse correlator decays as a power law with an exponent
$\eta \simeq 0.12\!-\!0.32$, which depends only weakly on the gasket generation and hardly at all on how distance is exactly defined. Thus, quantum fluctuations leave quasi-long-range order
(QLRO) intact. We then discuss what this implies for the
finite-temperature physics of realistic structures, which are finite by construction and for which the thermodynamic limit is of limited relevance. This uncovers an anticipated design trade-off where increasing gasket generation enhances the pairing amplitude, while phase coherence favors smaller systems.

\section{Model and mapping}

\subsection{From a Josephson network to the quantum XY model}
We start by considering a network of superconducting islands coupled by weak Josephson links, forming a graph.
%connected by Josephson coupling to the quantum XY model. 
%Consider a set of superconducting islands .
When the pairing gap on each island, labeled by $i$, is large compared with both the island charging energy $E_C$
and the Josephson coupling $E_J$, the magnitude of the pairing amplitude is fixed and the
low-energy dynamics is carried by the superconducting phases $\theta_i$, described by the quantum
phase (rotor) model \cite{dutta2001phase, medvidovic2023variational}
\begin{equation}
    H_{\mathrm{rotor}} = H_C + H_J = \frac{E_C}{2}\sum_i \bigl(\hat n_i - n_g\bigr)^2
    - E_J \sum_{\langle i,j\rangle}\cos(\theta_i-\theta_j),
    \label{eq:rotor}
\end{equation}
where $\hat n_i$ counts
the Cooper pairs on island $i$, and $n_g$ is the offset charge set by an external gate potential.
We work in the charge eigenbasis, $\hat n_i|n\rangle_i = n|n\rangle_i$ with
$n \in \mathbb{Z}$, in which $H_C$ is diagonal and the phase enters through the
unitary shift $e^{i\theta_i} = \sum_n |n+1\rangle\langle n|_i$. The Josephson term
\begin{equation}
    \cos(\theta_i-\theta_j) = \tfrac{1}{2}\bigl(e^{i\theta_i}e^{-i\theta_j}
    + \mathrm{h.c.}\bigr)
\end{equation}
therefore transfers a single Cooper pair between neighboring islands.
For convenience we take $n_g = 1/2$, at which the two charge states $n_i = 0, 1$ are
degenerate, with $E(0) = E(1) = E_C/8$, while every other charge state lies at least
$E_C$ higher. When $E_C \gg E_J$, the Josephson term is too weak to populate those
higher energy states and the low-energy Hilbert space is spanned by the on-site states
$\{|0\rangle_i, |1\rangle_i\}$, where $|1\rangle_i$ carries one Cooper pair of
charge $2e$ relative to $|0\rangle_i$. Each island thus holds at most one pair,
although the ground state within this space is not simply a product of on-site states.

Next, let $P_i = |0\rangle\langle 0|_i + |1\rangle\langle 1|_i$ project onto these two
states, with $P = \prod_i P_i$. Restricted to the image of $P$, the charging term $H_C$ is
the constant $N E_C/8$, which we henceforth drop. Acting on the low-energy space, $e^{i\theta_i}$ takes $|0\rangle_i \to |1\rangle_i$
and $|1\rangle_i \to |2\rangle_i$, but the latter is removed by the projection,
leaving
\begin{equation}
    P_i\,e^{i\theta_i}P_i = |1\rangle\langle 0|_i \equiv b_i^{\dagger},
    \qquad
    P_i\,e^{-i\theta_i}P_i = |0\rangle\langle 1|_i \equiv b_i .
\end{equation}
For $i \neq j$, the two exponentials act on distinct tensor factors, so the
projection of their product factorizes exactly,
$P\,e^{i\theta_i}e^{-i\theta_j}P = b_i^{\dagger}b_j$, and thus
\begin{equation}
    P\cos(\theta_i-\theta_j)P
    = \tfrac{1}{2}\bigl(b_i^{\dagger}b_j + b_j^{\dagger}b_i\bigr).
\end{equation}
The operators $b_i,b_i^\dagger$ obey $(b_i^{\dagger})^2 = 0$ and $\{b_i, b_i^{\dagger}\} = 1$ on a
given site $i$, while commuting on different sites. They are thus hard-core bosons with
$b_i^{\dagger}b_i = |1\rangle\langle 1|_i$ counting the Cooper pairs on island $i$. Transferring a
pair between an occupied and an empty island stays within the low-energy space, so
the projected Josephson term is nonzero and gives
\begin{equation}
    H_{\mathrm{hc}} = -\frac{E_J}{2}\sum_{\langle i,j\rangle}
    \bigl(b_i^{\dagger}b_j + b_j^{\dagger}b_i\bigr).
    \label{eq:hardcore}
\end{equation}
Processes that leave the low-energy space, such as moving a pair onto an already occupied
island, cost $E_C$ and also generate a nearest-neighbor density-density interaction of order $E_J^2/E_C$, negligible in the limit
considered here. 
For $|n_g - 1/2| < 1$, but where the same two charge states still remain
lowest, detuning $n_g$ adds a chemical potential $-\mu\sum_i b_i^{\dagger}b_i$ with
$\mu = (E_C/2)(1 - 2n_g)$. This term commutes with Eq.~\eqref{eq:hardcore} and thus at most shifts which pair-number sector lies lowest, and therefore we drop it.

The next step is to identify that hard-core bosons are equivalent to spin-1/2 operators. Under the Matsubara-Matsuda
identification~\cite{matsubara1956lattice}
\begin{equation}
    S_i^{+} = b_i^{\dagger}, \qquad S_i^{-} = b_i, \qquad
    S_i^{z} = \hat n_i - \tfrac{1}{2},
\end{equation}
the hard-core constraint becomes the two-dimensionality of the on-site Hilbert space of spin operators, $n_g = 1/2$ corresponds to $\sum_i S_i^z = 0$, and the pair hopping turns into a transverse exchange,
\begin{equation}
    b_i^{\dagger}b_j + b_j^{\dagger}b_i = S_i^{+}S_j^{-} + S_i^{-}S_j^{+}
    = 2\bigl(S_i^{x}S_j^{x} + S_i^{y}S_j^{y}\bigr).
\end{equation}
All this results in Eq.~\eqref{eq:hardcore} transforming into the spin-1/2 XY ferromagnet
\begin{equation}
    H = -J\sum_{\langle i,j\rangle}\bigl(S_i^{x}S_j^{x} + S_i^{y}S_j^{y}\bigr),
    \qquad J = E_J > 0,
    \label{eq:H_xy}
\end{equation}
with $S_i^{x,y} = \tfrac{1}{2}\sigma_i^{x,y}$.

We here stress that the word ``island'' carries no commitment to a
particular length scale. In our earlier work on fractal
superconductors~\cite{iliasov2026strong,canyellas2026extended}, the lattice was
atomistic, with each island being a single atom of an artificial crystal of the kind
assembled by STM manipulation~\cite{kempkes2019design} or molecular
self-assembly~\cite{shang2015assembling}.%, and the pairing was treated microscopically within BdG theory. 
The Josephson description of
Eq.~\eqref{eq:rotor} is also directly relevant to mesoscopic arrays of lithographic
junctions~\cite{gordon1986fractal,meyer1991vortex,meyer2002dimensional}. An effective Josephson lattice has even been derived
directly from the Hubbard model for the cuprates, with the coupling obtained from
microscopic correlation functions~\cite{harland2019josephson}. What is required is simply that the energy cost of adding a Cooper pair to a site exceeds the coupling between sites, and this holds at the atomic scale as much as for micron-sized grains. In a single orbital, it is the Pauli principle forbidding a second pair outright, so the effective charging energy is naturally large. The relevant parameters 
are thus the ratio $E_J/E_C$ and also the connectivity of the graph, not the physical size of an island. Equation~\eqref{eq:H_xy} therefore describes the phase sector of a fractal
superconductor at both extremes, both an atomistic lattice and a lithographic array, and
the ratio $E_J/E_C$ is what interpolates between the nearly classical and the
strongly quantum regimes.

For the spin-1/2 XY ferromagnet in Eq.~\eqref{eq:H_xy}, the in-plane magnetization $\langle S^x_i\rangle$ translates to the condensate order parameter
$\langle b_i\rangle$ in Eq.~\eqref{eq:hardcore}, while the transverse correlator
\begin{equation}
  \mathcal{C}_{xy}(i,j)
  = \bigl\langle \sigma^x_i\sigma^x_j + \sigma^y_i\sigma^y_j\bigr\rangle
  \;\propto\; \bigl\langle b^\dagger_i b_j + \mathrm{h.c.}\bigr\rangle
  \label{eq:tpc}
\end{equation}
is the off-diagonal one-body density matrix. The long-distance behavior of the latter is the standard diagnostic: a nonzero limit means off-diagonal long-range order, algebraic decay means a quasi-condensate, while exponential decay is an insulator. As Eq.~\eqref{eq:H_xy} is
invariant under global $U(1)$ rotations about the $z$ axis, $\langle S^x_i\rangle$
vanishes identically in any finite system. The correlator, or equivalently the
squared in-plane magnetization
\begin{equation}
  \bigl\langle M_\perp^2\bigr\rangle
  = \sum_{i \neq j}\bigl\langle \sigma^x_i\sigma^x_j
    + \sigma^y_i\sigma^y_j\bigr\rangle
  = \sum_{i \neq j} \mathcal{C}_{xy}(i,j),
  \label{eq:mperp}
\end{equation}
is therefore the appropriate order parameter~\cite{sandvik2010computational}.

\subsection{Geometries and distance measures}

We consider the triangular Sierpi\'nski gasket of generation $G$, which has
$N = 3(3^G+1)/2$ sites ($N = 6, 15, 42, 123, 366$ for $G = 1,\dots,5$) and
$3^{G+1}$ bonds, alongside its honeycomb variant, which has $N = (5\cdot 3^G+3)/2$
sites ($N = 9, 24, 69, 204, 609$) and an identical bond count, see Fig.~\ref{fig:sierpinski}.
To measure the correlator in Eq.~\eqref{eq:mperp}, we need a notion of distance. We therefore measure the separation
between sites $i$ and $j$ in two ways: by the graph distance $d_{ij}$, i.e.~the
number of bonds along the shortest path, and by the Euclidean distance $r_{ij}$ in
the embedding plane. In each case, we average over all pairs at a
given separation,
\begin{equation}
\label{eq:corr}
  C_{xy}(d) = \frac{1}{\mathcal{N}_d}\sum_{d_{ij} = d}\mathcal{C}_{xy}(i,j),
\end{equation}
where $\mathcal{N}_d$ is the number of such pairs, and analogously for, $C_{xy}(r)$ with $r_{ij}$ in place of $d_{ij}$. In addition, we consider pairs lying on the outer boundary between the extremal corners, where
the two measures coincide. These are the longest paths in the structure and the
ones most exposed to disconnection, so comparing their decay with the bulk tests
whether the algebraic behavior is a property of the state as a whole.

\subsection{Vision-transformer ansatz and variational optimization}
Beyond the smallest generations of the considered fractal structures, the Hilbert space is far out of reach of exact methods, and so we turn to a variational approach.
We therefore parametrize the many-body ground state wave function by a vision-transformer neural quantum
state~\cite{carleo2017solving,vaswani2017attention,viteritti2023transformer},
optimized by variational Monte Carlo within NetKet~\cite{vicentini2022netket}. The
gaskets have no translational symmetry to exploit and no natural one-dimensional
ordering, so neither symmetric convolutional architectures nor matrix-product
states apply directly, whereas the attention mechanism handles an unstructured
graph without imposing locality. Working with the wave function itself also gives access to all-to-all correlators
and to fidelities between independently optimized states, which we use in
Appendix~\ref{app:fidelity} to assess the reproducibility of the optimization.

\begin{figure}
    \centering
    \includegraphics[width=\linewidth]{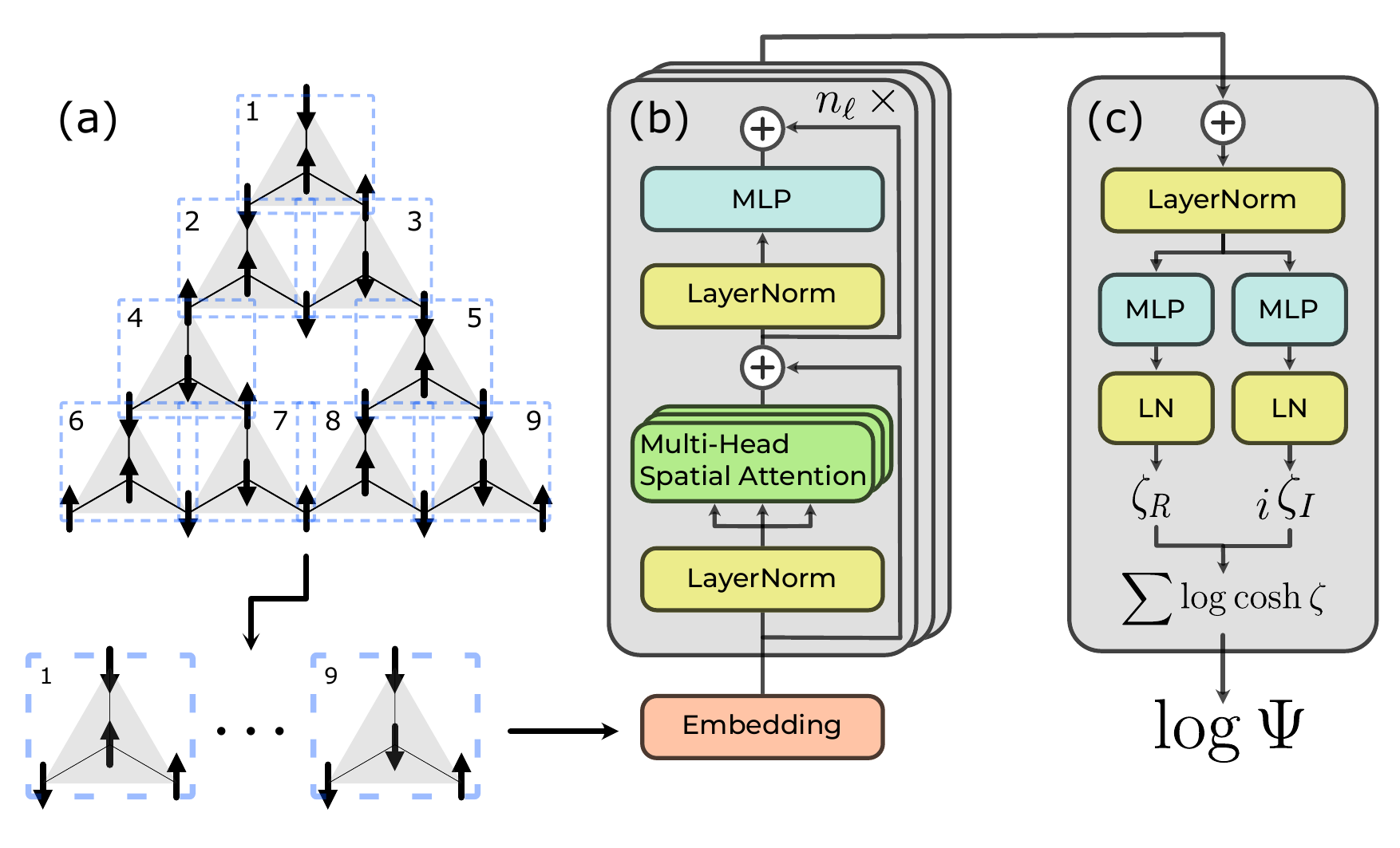}
    \caption{Vision-transformer architecture. (a) The dashed blue boxes indicate the patches into which the fractal is tokenized. The arrows denote some spin configuration $\boldsymbol{\sigma}$. (b) The tokens are processed by $n_\ell$ transformer encoder layers. (c) The output layer maps the result onto the log-amplitude $\log\Psi(\boldsymbol{\sigma})$.}
    \label{fig:architecture}
\end{figure}

The ansatz consists of three stages, shown in Fig.~\ref{fig:architecture}: (a) the embedding of the spin configuration into a sequence of patch representations, (b) a stack of transformer encoder layers, and (c) an output layer that maps the result onto the complex logarithm of the wave function, $\log\Psi(\boldsymbol{\sigma})$~\cite{carleo2017solving}, for some spin configuration $\boldsymbol{\sigma}$. The patches in Fig.~\ref{fig:architecture}(a) are the smallest building blocks of the gasket, indicated by the dashed regions and, unlike in conventional ViT implementations~\cite{viteritti2023transformer,viteritti2025transformer}, here they are allowed to overlap. Each patch is linearly projected into a $D$-dimensional embedding space. The resulting sequence of patch embeddings, $\boldsymbol{x} = (\boldsymbol{x}_1,\dots,\boldsymbol{x}_{n_p})$, one $D$-dimensional token per patch, is the input to the encoder.

Each encoder layer in Fig.~\ref{fig:architecture}(b) consists of a pre-normalized~\cite{pmlr-v119-xiong20b} factored
multi-head spatial attention block, followed by a feed-forward network (FFN), with
residual connections~\cite{7780459} around both; we omit the layer index. Here, after
layer normalization (LN), the patch representations are passed through a learnable value
projection and split among $n_h$ heads, giving
$\boldsymbol{v}^{(h)}_j \in \mathbb{R}^{D/n_h}$. The output of head $h$ for patch
$i$ is
\begin{equation}
    \boldsymbol{A}^{(h)}_i = \sum_j \alpha^{(h)}_{ij}\, w_{ij}\, \boldsymbol{v}^{(h)}_j ,
\end{equation}
where the attention weights $\alpha^{(h)}_{ij}$ are learnable parameters, and
$w_{ij}$ is a distance-dependent kernel, which softly biases the attention weights~\cite{viteritti2026approaching}, 
\begin{equation}
    w_{ij} = \frac{(1 + r_{ij})^{-\gamma}}{\sum_k (1 + r_{ik})^{-\gamma}} ,
\end{equation}
with $r_{ij}$ the Euclidean distance between the centers of patches $i$ and $j$.
The exponent $\gamma$ is learnable and shared by all heads within a layer, while
each head has its own attention matrix $\alpha^{(h)}$ and acts on its own
$D/n_h$-dimensional value subspace, so that different heads can learn different
spatial correlations.
The head outputs are then concatenated into a single vector $\boldsymbol{A}_i$ per patch
and linearly projected, which mixes information between heads. The input
representation is then added back through the residual connection. The second
sub-block is a position-wise FFN, a two-layer multilayer perceptron (MLP) with a
Gaussian Error Linear Unit (GELU) activation. Here, after the second layer normalization
the hidden dimension is expanded from $D$ to $4D$ and projected back to $D$, and a
second residual connection gives the output of that layer. This is repeated for
$n_l$ encoder layers, the output of layer $\ell$ serving as the input of layer
$\ell + 1$.

After the last encoder layer, 
the patch representations are summed and
layer-normalized into a single global vector, shown in Fig.~\ref{fig:architecture}(c). This vector is passed through two
independent output branches, each a MLP followed by its own layer
normalization, which produce the real and imaginary feature vectors $\zeta_R$ and
$\zeta_I$, combined into $\zeta = \zeta_R + i\zeta_I$. The logarithm of the
variational wave function is then finally given as
\begin{equation}
    \log\Psi(\boldsymbol{\sigma}) = \sum_{i=1}^{D} \log\cosh\zeta_i .
\end{equation}

The network parameters are optimized by variational Monte Carlo
(VMC)~\cite{becca2017quantum}, with spin configurations sampled from
$|\Psi(\boldsymbol{\sigma})|^2$. The energy is minimized by stochastic
reconfiguration (SR)~\cite{PhysRevB.71.241103} in its minSR
formulation~\cite{chen2024empowering}, which makes SR affordable at large
parameter counts. We use $N_s = 2^{13}$ Monte Carlo samples spread over as many
Markov chains, combined with the SPRING momentum
scheme~\cite{goldshlager2024kaczmarz} with a momentum of $0.8$. The optimization
of the variational state is performed in two stages. In the first stage the total
magnetization is not fixed: configurations are sampled with local Metropolis
updates, and the state is optimized for 750 iterations with an initial learning
rate of $10^{-2}$ with exponential decay and a diagonal shift of
$5\times 10^{-4}$. In the second stage the total magnetization is fixed to
$S^z = 0$ for an even number of sites and to $S^z = 1/2$ for an odd number, and
configurations are sampled with Metropolis exchange updates, which preserve $S^z$.
The network is warm-started from the converged states of the first stage and
optimized for a further 400 iterations. The resulting state is then symmetrized
under the $C_{3v}$ point group of the gasket by quantum-number
projection~\cite{Nomura_2021} and optimized for another 20 iterations. For seeds
whose variance deviates strongly from the others, the second stage is repeated. Hyperparameters tuned for one generation or gasket type need not transfer to another, so we select the network and learning-rate hyperparameters by a grid search minimizing the average energy variance at $G = 3$ and $G = 4$.

Convergence of the variational state towards the ground state is assessed by
monitoring the energy and the energy variance per site, which we combine into the
V-score~\cite{wu2024variational}
\begin{equation}
  V = \frac{N\,\mathrm{Var}(E)}{(E - E_\infty)^2},
\end{equation}
where $E_\infty = \mathrm{Tr}\,H/\dim\mathcal{H}$ is the infinite-temperature
energy, which vanishes for the XY Hamiltonian.
Convergence of the energy at an early stage does not guarantee that the variational quantum state is free from excited-state contributions. A decreasing energy variance per site indicates that the wavefunction is approaching an eigenstate of the Hamiltonian. The simultaneous convergence of these two quantities, with further corroboration of the high fidelity between different optimized states depicted in Appendix~\ref{app:fidelity}, provides strong evidence that the variational quantum state approaches the ground state.
%Figure~\ref{fig:convergence} shows the optimization for each generation. The energy per site (a,b) saturates early in the optimization. A more sensitive measure of convergence is the variance of the variational energy (bottom row). We find that the variance of the energy of the ground state with the system size, indicating that larger gaskets are increasingly more difficult to accurately represent with the same network, as is expected. To make sure that the optimization nevertheless reaches the same state, at $G = 4$ and $G = 5$ we optimize five independently initialized states and compute their pairwise fidelities in Appendix~\ref{app:fidelity}. These show excellent overlap, confirming that all runs converge to the same ground state.

\begin{figure}
    \centering
    \includegraphics[width=\linewidth]{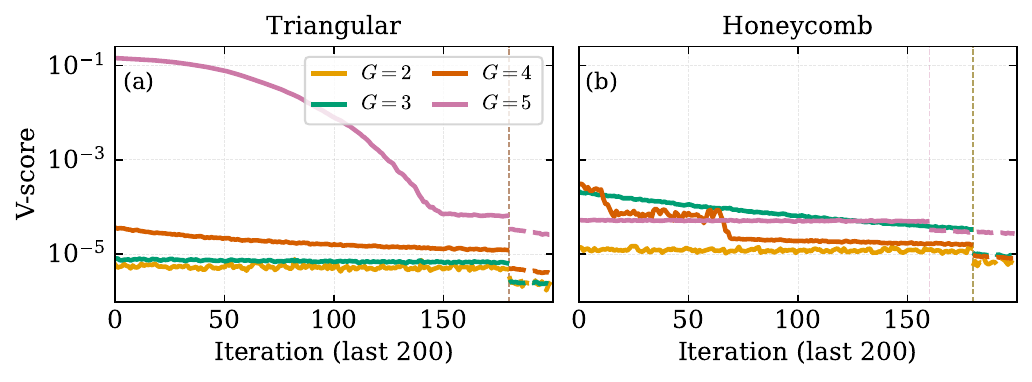}
    \caption{V-score over the last 200 iterations for the (a) triangular and (b) honeycomb
Sierpi\'nski gaskets. Dashed vertical lines mark the iteration from which the
$C_{3v}$ symmetry projection is enforced. V-scores for $G=1$ are omitted, as they
repeatedly drop below $\mathcal{O}(10^{-5})$ and recover during the VMC iterations.}
    \label{fig:convergence}
\end{figure}
\section{Results}

\begin{figure}
    \centering
    \includegraphics[width=\linewidth]{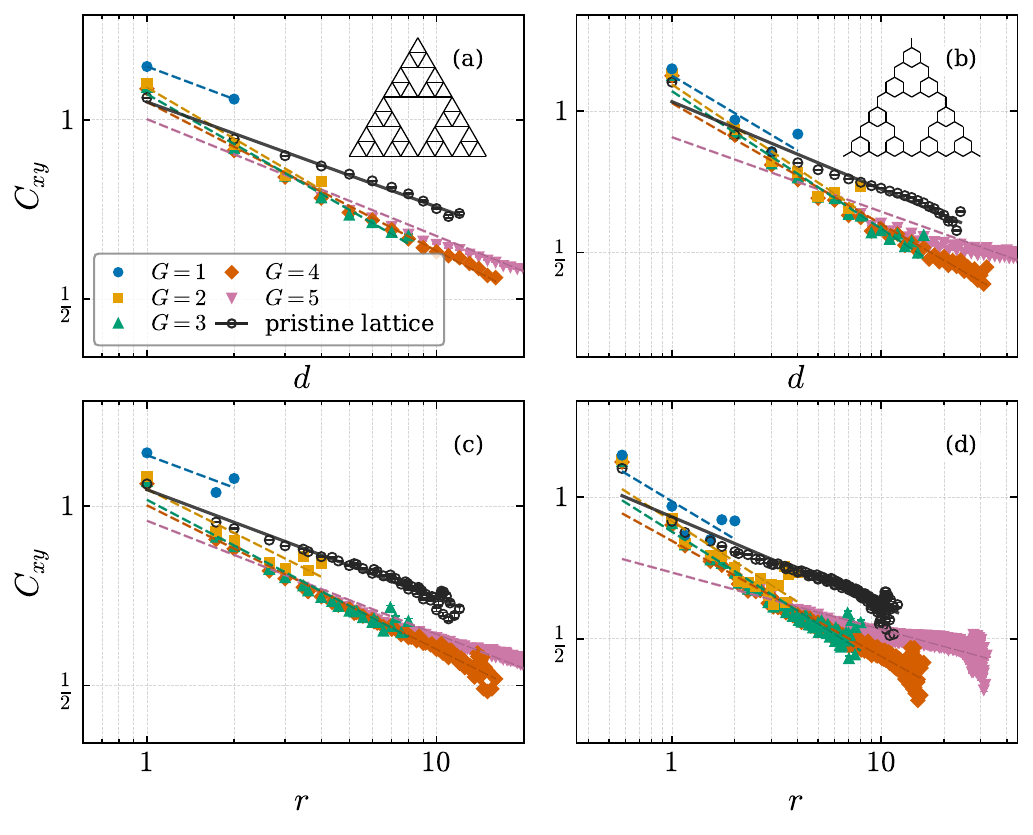}
    \caption{Two-point shell-averaged correlators $C_{xy}$ for the triangular Sierpi\'nski
    gasket (a,c) and the honeycomb Sierpi\'nski gasket (b,d) as a function of
    graph distance $d$ (a,b) and Euclidean distance $r$ (c,d), for different generations $G$ of each fractal. Straight lines on the double-logarithmic scale
    correspond to algebraic decay fits to Eq.~\eqref{eq:powerlaw}.}
    \label{fig:correlator}
\end{figure}

Using the ViT NQS to calculate the ground state, we extract the transverse two-point shell-averaged correlator in Eq.~\eqref{eq:corr}, averaged over 5 independently optimized variational states, to understand phase coherence in the rotor model. We plot the result in 
Fig.~\ref{fig:correlator} for both the triangular (a,c) and honeycomb (b,d) Sierpi\'nski gaskets. For each gasket, we report both the correlator as a function of network distance $d$ and Euclidean distance $r$. For comparison, we also include the pristine lattice, which is not a fractal.
On a double-logarithmic scale, all data fall on straight lines over the
full accessible range of distances, for every generation of both fractals and for both metrics. 
This leads to the fits
\begin{equation}
  C_{xy}(r) = A\,r^{-\eta}
  \label{eq:powerlaw}
\end{equation}
with the exponents in Table~\ref{tab:coefs}. Throughout, $\eta$ lies between
$0.12$ and $0.32$. Thus, finite and positive, so there is no true long-range order, but still well below
unity, so the decay is slow and the state is therefore a quasi-condensate. Notably, the two metrics
give the same exponent to within a few percent, and so do boundary pairs, which
follow the longest paths and are the most exposed to disconnection. The algebraic
decay is thus a property of the state as a whole rather than of a particular
projection of the geometry. The amplitude $A$ stays close to unity in the graph
metric but falls with $G$ in the Euclidean one, as expected from the growing
mismatch between chemical and embedding distance as the gasket is refined.
Note that five generations span about two decades in linear
size. This is enough to establish the algebraic character of the decay and to
bracket the exponent. However, it is not enough to extract $\eta$ in the thermodynamic limit
or to interpret its mild drift with $G$.
If we instead attempt an exponential fit, $C_{xy} = B\,e^{-r/\xi}$, it performs worse throughout and returns correlation lengths comparable to the system size, as expected for an exponential fit to algebraic data.

\begin{table}[htbp]
    \centering
    \caption{Fit parameters of Eq.~\eqref{eq:powerlaw} for the triangular and honeycomb Sierpi\'nski gaskets across three different distance measures.}
    \label{tab:coefs}
    \begin{ruledtabular}
    \begin{tabular}{lcccccc}
      & \multicolumn{2}{c}{graph}
      & \multicolumn{2}{c}{Euclidean}
      & \multicolumn{2}{c}{boundary}\\
      & $A$ & $\eta$ & $A$ & $\eta$ & $A$ & $\eta$\\
      \hline
      \multicolumn{7}{l}{Triangular gasket}\\
      $G=1$ & 1.23 & 0.182 & 1.22  & 0.181 & 1.29 & 0.210 \\
      $G=2$ & 1.13 & 0.285 & 1.07  & 0.244 & 1.15 & 0.292 \\
      $G=3$ & 1.10 & 0.278 & 1.02  & 0.255 & 1.10 & 0.290 \\
      $G=4$ & 1.07 & 0.250 & 1.00  & 0.242 & 1.07 & 0.278 \\
      $G=5$ & 1.00 & 0.196 & 0.945 & 0.190 & 1.03 & 0.232 \\
      Pristine & 1.07 & 0.176 & 1.07 & 0.182 & 1.09 & 0.210 \\
      \hline
      \multicolumn{7}{l}{Honeycomb gasket}\\
      $G=1$ & 1.19 & 0.265 & 0.980 & 0.268 & 1.21 & 0.290 \\
      $G=2$ & 1.14 & 0.298 & 0.890 & 0.286 & 1.15 & 0.315 \\
      $G=3$ & 1.10 & 0.292 & 0.845 & 0.280 & 1.10 & 0.314 \\
      $G=4$ & 1.04 & 0.257 & 0.808 & 0.245 & 1.05 & 0.286 \\
      $G=5$ & 0.880 & 0.158 & 0.692 & 0.121 & 0.895 & 0.180 \\
      Pristine & 1.05 & 0.186 & 0.909 & 0.189 & 1.05 & 0.216 \\

    \end{tabular}
    \end{ruledtabular}
\end{table}

To summarize, within the stated numerical limits, our conclusion is unambiguous and holds for both geometries,
all generations, and both metrics: the ground state of the quantum XY model on the
Sierpi\'nski gasket supports quasi-long-range order. In terms of superconductivity, this means that the extreme quantum limit of a fractal Josephson network is a quasi-condensate.

\section{Discussion of finite structures and finite temperature}
Having established quasi-long-range order in the ferromagnetic spin-1/2 XY model for finite fractals, we come back to the actual question of superconducting phase coherence at finite temperatures relevant for experimentally accessible fractal superconductors.
%What one should ask of a fractal superconductor deserves a closer look. 
First of all, the classical results cited in the introduction concern the thermodynamic limit. They follow the helicity modulus (the response of the free energy to a twist of the
boundary phases, that is the phase stiffness of the classical model) as the gasket
generation is increased without bound~\cite{vallat1991xy,mitrovic2010monte,przedborski2011absence}. For a
bulk material, this limit of ever-increasing gasket generations is the natural limit. However, for current experimental structures it
is not: an STM-assembled fractal contains dozens to a few hundred atoms, a lithographic gasket array is fabricated at low generation, such that the sample is always finite by construction. Thus the relevant question must be how a gasket of finite generation $G$ behaves at the temperatures and length scales at which it is operated.

Second, one might be tempted to read algebraic correlations with $\eta \simeq 1/4$ found in this work as evidence of BKT physics. Here $\eta = 1/4$ follows from $\eta = k_B T / 2\pi K_R$ at the universal jump of the helicity modulus, $K_R = 2/\pi$~\cite{kosterlitz1973ordering,%
nelson1977universal,komura2012large}, and therefore requires a finite
$T_{\mathrm{BKT}}$. However, on a finitely ramified fractal the classical transition temperature vanishes, and the same relation gives $\eta \to 0$. This is consistent with the classical ground state, which on any connected graph is fully aligned and has true long-range order rather than algebraic decay. The algebraic correlations we find are thus of quantum origin, and the thermal mechanism that fixes $1/4$ cannot produce them. Our exponents are instead set by the lattice geometry and by the quantum fluctuations of the phase, and their proximity to $1/4$ does not, by itself, indicate an underlying connection to BKT criticality.

Thus, posed this way, the problem at finite temperatures in finite fractals separates into a quantum part and a classical part.
The quantum part is settled by this work: at $T = 0$, and at the maximal
strength of quantum phase fluctuations compatible with superconductivity, the
finite gasket carries algebraic off-diagonal correlations with an exponent well
below unity. This result complements the mean-field results of
Ref.~\cite{iliasov2026strong}, where the superfluid stiffness of these gaskets was
found to be finite and comparable to that of the regular lattice. Since our limit of $E_C \gg E_J$ is the extreme quantum limit of the rotor model, this is the least favorable case for phase coherence. At any finite $E_C/E_J$ the phase fluctuations are weaker and the classical XY limit is even recovered as $E_C/E_J \to 0$. Quantum fluctuations of the phase, then, are not what limits the existence of a finite fractal superconductor.

The classical part of the problem is the finite-size behavior at $T > 0$, which we intend to
address in depth in future work, though the outline of the answer is already clear. A gasket of generation $G$ has a linear size $2^G$. Since the vortex energy is
bounded, free vortices are present at any $T > 0$ with a density that sets a mean
spacing $\ell_v(T)$ between them, and the phase stays coherent over distances
shorter than $\ell_v$. A finite structure therefore retains a superconducting
response as long as $\ell_v(T) \gtrsim 2^G$, which defines a crossover temperature
that depends on $G$, but that the thermodynamic-limit analysis integrates out. The existing experiments are consistent with this picture of existing phase coherence. In
periodic arrays of gaskets of order $G$ joined at their vertices, where the
geometry is fractal below the gasket size and Euclidean above it, a BKT transition
is recovered and shows up as a sharp drop of the inverse sheet magnetoinductance,
with transition temperatures following $\tau_{c,G} = (3/5)^G \tau_{c,0}$, where
$\tau_{c,0} = k_B T/E_J \simeq 1.5$ for a regular triangular array of
the same junctions~\cite{meyer2002dimensional}. Therefore, higher generations of fractals show diminishing phase coherence, though it remains measurable in smaller systems.

This illustrates an anticipated design trade-off. Increasing the gasket generation sharpens the
spectral clustering that enhances the pairing amplitude~\cite{iliasov2026strong},
while the phase stiffness, and with it phase coherence, moves the other way; in the periodic arrays of
Ref.~\cite{meyer2002dimensional} it even falls by a factor of $3/5$ per generation. Since
the actual superconducting transition temperature $T_c$ is bounded by the smaller of the two scales~\cite{emery1995importance}, the
optimum necessarily lies at a finite, and presumably rather low, generation. The figure of merit can thus be thought of as the product of amplitude and stiffness rather than the gap enhancement alone. Note however that how strongly the stiffness of a single gasket falls with $G$ is a
separate question from the array measurement, and a harder one. The helicity
modulus is defined through a twist imposed across the system, which the periodic
arrays of Ref.~\cite{meyer2002dimensional} realize but an isolated gasket with open
boundaries does not. A response can still be defined, either by threading flux
through the loops of the gasket or from the current-current correlator of
Eq.~\eqref{eq:H_xy}, and our variational states give access to both; which of these
is the appropriate finite-size analogue of the stiffness we leave for future work. %Note however, that how strongly the stiffness of a single gasket falls with $G$ is a separate question from the array measurement, and answering it requires the stiffness of the ground state of Eq.~\eqref{eq:H_xy} -- which is $J$ reduced by geometry and by quantum fluctuations, with the mean-field value found in Ref.~\cite{iliasov2026strong} setting the scale of $J$ itself. It can be obtained from the response to a twist of the boundary phases, which our variational states make accessible, and we leave it for future work.

\section{Conclusion}

We compute the ground state of the ferromagnetic spin-$1/2$ XY model, which is the phase sector of a Josephson network in the limit of at most one Cooper pair per island, i.e.~the extreme quantum limit, on both
triangular and honeycomb Sierpi\'nski gaskets up to generation $G = 5$, using
vision-transformer neural quantum states. This architecture reaches system sizes
of several hundred sites on a lattice with no translational symmetry and no
natural one-dimensional ordering, where exact methods can at most handle a few dozen sites.

We find that the transverse correlations decay algebraically with $\eta \simeq 0.12\!-\!0.32$, the same for different metrics and for both geometries, and the exponent drifts only weakly
with generation. Quantum phase fluctuations therefore do not destroy
quasi-long-range order on a finitely ramified fractal, not even when they are
at their strongest. Together with the recent mean-field prediction of an enhanced pairing
amplitude on the same geometry \cite{iliasov2026strong, canyellas2026extended}, this locates the only remaining obstacle to fractal superconductivity in the classical, finite-temperature sector. Notably, the theorems that forbid ordering on an infinite gasket have limited bearing on structures of a few hundred sites, and the careful finite-size classical analysis is the natural next step. Our reasoning suggests a trade-off in gasket size, with larger gasket generations yielding a higher pair amplitude but lower phase coherence.
Two clear questions still remain. The phase stiffness of the ground state would turn the trade-off between pairing
amplitude and phase coherence into a quantitative statement about the optimal
gasket generation, but as discussed above, the standard definition through an
imposed twist does not carry over to an isolated fractal with open boundaries, and
settling which response takes its place is part of the problem. A second open
question is the rotor model at finite $E_C/E_J$, which interpolates between the
case treated here and the classical limit; we expect coherence to survive
throughout, but the crossover has not been mapped.

%We have computed the ground state of the spin-$1/2$ XY model, which is the phase sector of a Josephson network in the extreme quantum limit, on triangular and honeycomb Sierpi\'nski gaskets up to generation $G=5$, using vision-transformer neural quantum states. Transverse correlations decay algebraically with $\eta\simeq0.2\!-\!0.3$, robustly across geometries, generations, and definitions of distance. Quantum phase fluctuations therefore do not destroy quasi-long-range order on a finitely ramified fractal, not even in the extreme quantum limit where they are strongest.

%Together with the mean-field prediction of an enhanced pairing amplitude on the same geometry, this places the remaining obstacles to fractal superconductivity in the classical, finite-temperature sector, and at finite system size, where the thermodynamic-limit theorems that forbid ordering on an infinite gasket have limited bearing on structures of a few hundred sites. We regard the finite-size classical analysis as the natural next step.

\begin{acknowledgments}
R.C., M.I.K., and A.A.B. were supported by the research program “Materials for the Quantum Age” (QuMat). This program (registration number 024.005.006) is part of the Gravitation program financed by
the Dutch Ministry of Education, Culture and Science (OCW). A.A.B. was also supported by NWO grant OCENW.M.23.044.
J.v.P~and A.M.B.S.~acknowledge funding from the European Research Council (ERC) under the European Union’s Horizon 2020 research and innovation programme (ERC-2022-CoG, Grant agreement No. 101087096). Views and opinions expressed are, however, those of the author(s) only and do not necessarily reflect those of the European Union or the European Research Council Executive Agency. Neither the European Union nor the granting authority can be held responsible for them. 
The computations were enabled by resources in project UPPMAX 2026/1-2 provided
by Uppsala University at UPPMAX, as well as the National Academic
Infrastructure for Supercomputing in Sweden (NAISS), partially funded by the
Swedish Research Council through grant agreement no.~2022-06725. 
\end{acknowledgments}

\section*{Data availability}
The code and data required to reproduce the results of this work are openly available \cite{FractalXY_code}.

\bibliography{ref_updated}

\vspace*{1.25\baselineskip} % extra visual gap before the appendix (without forcing a page/column break)
\appendix
\section{Fidelity analysis}\label{app:fidelity}
\begin{figure}[H]
    \centering
    \includegraphics[width=\linewidth]{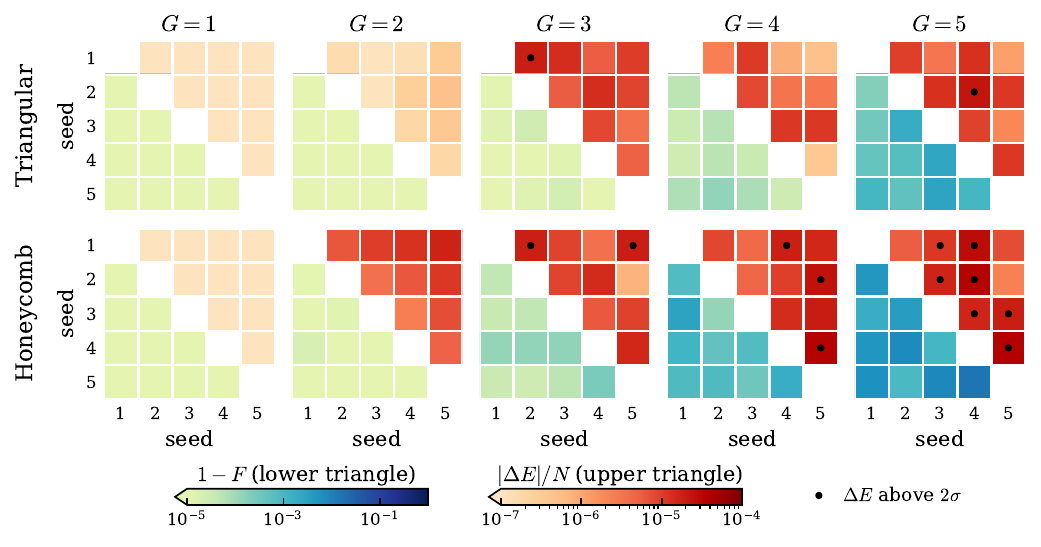}
    \caption{The lower triangle shows the pairwise infidelities $1-F$, and the upper triangle
the absolute energy difference per site, between independently optimized runs for
the triangular (top row) and honeycomb (bottom row) gaskets, showing that all runs
converge to the same state. Dots mark pairs whose energy difference exceeds
$2\sigma$.}
    \label{fig:fidelity}
\end{figure}
To check that the optimization converges to the same state regardless of
initialization, we performed five independent optimizations for each gasket
geometry at generations $G = 1$--$5$, differing only in the random seed used to
initialize the variational parameters. For each pair of optimized states we
computed the infidelity $1-F$, as implemented in NetKet~\cite{vicentini2022netket}.
For the smallest systems (triangular $G \le 2$, honeycomb $G = 1$) it was evaluated
exactly by summing over the full $S^z$ sector; for larger systems it was estimated
by Monte Carlo sampling. The results, shown in Fig.~\ref{fig:fidelity}, confirm
that different runs converge to the same state. Where energy differences are
statistically resolved, they remain below $10^{-4}$ per site and reflect small
differences in convergence rather than distinct states.

\end{document}